\documentclass[]{ase}

\usepackage{graphicx}
\usepackage{amssymb, amsmath, amsthm}
\usepackage{verbatim}
\usepackage{color}
\usepackage{microtype}

\usepackage[font=small,labelfont=sc]{caption}
\newenvironment{Figure}
  {\par\medskip\noindent\minipage[htb]{\linewidth}}
  {\endminipage\par\medskip}

\begin{document}

\begin{center}
{\huge Understanding the Predictive Power of Computational Mechanics and Echo State Networks in Social Media}



\setlength{\columnsep}{0pt}
\begin{multicols}{4}
{David Darmon}{\\}{\small Dept.\ of Mathematics\\
University of Maryland\\
College Park, MD 20740\\
ddarmon@math.umd.edu\\ \columnbreak}
{Jared Sylvester}{\\}{\small Dept.\ of Computer Science\\
University of Maryland\\
College Park, MD 20740\\
jsylvest@umd.edu\\ \columnbreak}
{Michelle Girvan}{\\}{\small Dept.\ of Physics\\
University of Maryland\\
College Park, MD 20740\\
girvan@umd.edu\\ \columnbreak}
{William Rand}{\\}{\small Ctr.\ for Complexity in Business\\
University of Maryland\\
College Park, MD 20740\\
wrand@rhsmith.umd.edu}\\
\end{multicols}
\end{center}

\begin{multicols*}{2}

\begin{abstract}
There is a large amount of interest in understanding users of social
media in order to predict their behavior in this space. Despite this interest, user predictability in social media is not well-understood. To examine this question, we consider a network of fifteen thousand users on Twitter over a seven week period. We apply two contrasting modeling paradigms: computational mechanics and echo state networks. Both methods attempt to model the behavior of users on the basis of their past behavior.  We demonstrate that the behavior of users on Twitter can be well-modeled as processes with self-feedback. We find that the two modeling approaches perform very similarly for most users, but that they differ in performance on a small subset of the users. By exploring the properties of these performance-differentiated users, we highlight the challenges faced in applying predictive models to dynamic social data.
\end{abstract}

\setcounter{collectmore}{20}
\raggedcolumns

\section{Introduction}








At the most abstract level, an individual using a social media service may be viewed as a computational agent~\cite{dedeo2012evidence}. The user receives inputs from their surroundings, combines those inputs in ways dependent on their own internal states, and produces an observed behavior or output. In the context of a microblogging platform such as Twitter, the inputs may be streams from other Twitter users, real world events, etc., and the observed behavior may be a tweet, mention, or retweet. From this computational perspective, the observed behavior of the user should give some indication of the \emph{types} of computations the user is doing, and as a result, an insight into viable behavioral models of that user on social media. Large amounts of observational data are key to this type of study. Social media has made such behavioral data available from massive numbers of people at a very fine temporal resolution.

As a first approximation to the computation performed by a user, we might consider only the user's own past behavior as possible inputs to determine their future behavior. From this perspective, the behavior of the user can be viewed as a point process with memory, where the only observations are the time points when social interactions occurred~\cite{perry2010point}. Such point process models, while very simple, have found great success in describing complicated dynamics in neural systems~\cite{rieke1999spikes}, and have recently been applied to social systems~\cite{ver2012information, cho2013latent}.

We propose extending this previous work by explicitly studying the \emph{predictive} capability of the point process models. That is, given observed behavior for the user, we seek a model that not only captures the dynamics of the user, but also is useful for predicting the future behavior of the user, given their past behavior. The rationale behind this approach is that if we are able to construct models that both reproduce the observed behavior and successfully predict future behavior, the models capture something about the computational aspects, in the sense outlined above, of the user.

Since in practice we never have access to all of a user's inputs, nor to their internal states, we cannot hope to construct a `true' model of a user's behavior. Instead, we construct approximate models. In particular, we consider two classes of approximate models: causal state models and echo state networks.

The causal state modeling approach, motivated by results from computational mechanics, assumes that every individual can initially be modeled as a biased coin, and then adds structure as necessary to capture patterns in the data.
It does this by expanding the number of states necessary to represent the underlying behavior of the agent.
Causal state models have been used successfully in various fields, including elucidating the computational structure of neural spike trains~\cite{haslinger2010computational}, uncovering topic correlations in social media~\cite{cointet2007intertemporal}, and improving named entity recognition in natural language processing~\cite{padro2005named}.
As opposed to the simple-to-complex approach used by causal state modeling, echo state networks start by assuming that agent behavior is the result of a complex set of internal states with intricate relationships to the output variables of interest, and then simplifies the weights on the relationships between the internal states and the output variables over time.
Echo state networks have proven useful in a number of different domains including wireless networking~\cite{Jaeger2004}, motor control~\cite{Salmen2005}, and grammar learning~\cite{Tong2007}.

Our motivation for considering these two models was twofold. First, they share a structural similarity in that they both utilize hidden states that influence behavior and incorporate past data when making future decisions. Second, they approach modeling from two different perspectives.
As mentioned, both representations have a notion of internal state, and the observation of past behavior moves the agent through the possible states.
It is the model of these dynamics through the states that makes it possible to use these methods to predict an individual's behavior.
Moreover, whereas computational mechanics seeks to construct the simplest model with the maximal predictive capability,  echo state networks relax down from very complicated dynamics until predictive ability is reached.
Due to this difference, we hypothesize that there are some users that will be easier to predict using a causal state modeling approach, and a different set of users that will be easier to predict using an echo state network approach.

In the rest of this paper, we explore this hypothesis.
We begin by describing the two approaches  we used and their relevant literature.
After this, we describe the data used to test the predictive ability of these methods, and the investigations that we carried out to evaluate this ability.
Finally, we conclude with limitations of the present work and future avenues of research.

\section{Methodology}

\subsection{Notation}

For each user, we consider only the relative times of their tweets with respect to a reference time. Denote these times by $\{ \tau_{j}\}_{j = 1}^{n}$. Let the reference start time be $t_{0}$ and the coarsening amount be $\Delta t$. From the tweet times, we can generate a binary time series $\{ X_{i}\}_{i = 1}^{T}$, where
\begin{align}
	X_{i} = \left\{ \begin{array}{cl}
		1 &: \text{$ \exists \tau_{j} \in [t_{0} + (i - 1) \Delta t, t_{0} + i \Delta t)$} \\
		0 &: \text{ otherwise}
	\end{array}\right. .
\end{align}
In words, $X_{i}$ is 1 if the user tweeted at least once in the time interval $[t_{0} + (i - 1) \Delta t, t_{0} + i \Delta t)$, and 0 otherwise. Because the recorded time of tweets is restricted to a 1-second resolution, a natural choice for $\Delta t$ is 1 second. However, due to limitations in the amount of data available we will coarsen the time series to less than this resolution. Thus, in this paper, we consider the behavior of the user as a point process, only considering the timing of the tweets, and discarding any informational content \emph{in} the tweet (sentiment, retweet, mention, etc.).

Once we have the user's behavior encoded in the sequence $\{ X_{i}\}_{i = 1}^{T}$, we wish to perform one-step ahead prediction based on the past behavior of the user. That is, for a time bin $[t_{0} + (i - 1) \Delta t, t_{0} + i \Delta t)$ indexed by $i$, we wish to predict $X_{i}$ given a finite history $X_{i - L}^{i - 1} = (X_{i-L}, \ldots, X_{i - 2}, X_{i - 1})$ of length $L$. This amounts to a problem in autoregression, where we seek a function $r$ from finite pasts to one-step ahead futures such that we predict $X_{i}$ using
\begin{align}
	\hat{X}_{i} = \operatorname*{arg\,max}_{x_{i} \in \{0, 1\}} r(x_{i}; x_{i - L}^{i - 1}).
\end{align}
If we assume that $\{ X_{i}\}_{i = 1}^{T}$ was generated by a stochastic process, the optimal choice of $r$ would be the conditional distribution
\begin{align}
	r(x_{i}; x_{i - L}^{i - 1}) = P(X_{i} = x_{i} | X_{i-L}^{i-1} = x_{i - L}^{i - 1}),
\end{align}
and the optimal prediction would be the $x_{i}$ that maximizes this conditional probability. If we further assume that $\{ X_{i}\}_{i = 1}^{T}$ is a conditionally stationary stochastic process~\cite{caires2003nonparametric}, the prediction function simplifies to 
\begin{align}
	r(x_{i}; x_{i - L}^{i - 1}) = P(X_{L} = x_{i} | X_{0}^{L-1} = x_{i - L}^{i - 1}),
\end{align}
independent of the time index $i$.

Because in practice we do not have the conditional distribution available, we consider two approaches to inferring the prediction function $r$: one from computational mechanics~\cite{shalizi2001computational} and the other from reservoir computing~\cite{Schrauwen2007}, specifically the echo state network~\cite{Jaeger2001}. These two methods for inferring $r$ differ dramatically in their implementations. Computational mechanics seeks to infer the simplest model that will capture the data generating process, while echo state networks generate a complex set of oscillations and attempt to find some combination of these that will map to the desired output.

\subsection{Computational Mechanics}

Computational mechanics proceeds from a state-space representation of the observed dynamics, with hidden states $\{S_{i}\}_{i = 1}^{T}$ determining the dynamics of the observed behavior $\{ X_{i}\}_{i = 1}^{T}$. The hidden state $S_{i}$ for a process, called the causal or predictive state, is the label corresponding to set of all pasts that have the same predictive distribution as the observed past $x_{i}.$ We call the mapping from pasts to labels $\epsilon$. Two pasts $x$ and $x'$ have the same label  $s_{i} = \epsilon(x) = \epsilon(x')$ if and only if
\begin{align}
 	P(X_{i} | X_{i-L}^{i - 1} = x) = P(X_{i} | X_{i-L}^{i - 1} = x')
\end{align}
as probability mass functions. Now, instead of considering $P(X_{i} | X_{i-L}^{i - 1} = x_{i - L}^{i - 1}),$ we consider the label for the past $s_{i} = \epsilon(x_{i - L}^{i - 1}),$ and use $P(X_{i} | S_{i} = s_{i})$. We then proceed with the prediction problem outlined above. The state $S_{i}$ (or equivalently the mapping $\epsilon$) is the unique minimally sufficient predictive statistic of the past for the future of the process. Because the hidden states $\{S_{i}\}_{i = 1}^{T}$ can be thought of as generating the observed behaviors $\{X_{i}\}_{i = 1}^{T}$, they are called the \emph{causal states} of the process. The resulting model is called an $\epsilon$-machine (after the statistic $\epsilon$) or a causal state model (after the causal state $S$).

Of course, in practice the conditional distribution $P(X_{i} | X_{i-L}^{i - 1} = x)$ is not known, and must be inferred from the data. Beyond the advantage of computational mechanics's state-space representation as a minimally sufficient predictive statistic, it also admits a way to infer the mapping $\epsilon$ directly from data. We will infer the model using the Causal State Splitting Reconstruction (CSSR) algorithm~\cite{CSSR-UAI-2004}. As the name CSSR implies, the estimate $\hat{\epsilon}$ is inferred by splitting states until a stopping criterion is met. The algorithm begins with a null model, where the data generating process is assumed to have a single causal state, corresponding to an IID process. It continues to split states (representing a finer partition of the set of all pasts) until the partition is next-step sufficient and recursively calculable. The resulting $\hat{\epsilon}$ and the estimated predictive distributions $\hat{P}(X_{i} | S_{i} = \hat{\epsilon}(x_{i - L}^{i - 1}))$ can then be used to estimate the prediction function, giving
\begin{align}
	\hat{r}_{\text{cm}}(x_{i}; x_{i-L}^{i - 1}) = \hat{P}(X_{i} = x_{i} | S_{i} = \hat{\epsilon}(x_{i - L}^{i - 1})).
\end{align}
We will refer to the estimated $\hat{\epsilon}$ and associated predictive distributions as the \emph{causal state model} for a user.

\subsection{Echo State Networks}

Neural networks can be divided into feed-forward and recurrent varieties. The former are easier to train but lack the capacity to build rich internal representations of temporal dynamics. In contrast, the latter are naturally suited to representing dynamic systems, but their learning algorithms are more computationally intensive and less stable. Echo state networks attempt to resolve this conflict
by using randomly selected, fixed weights to drive the recurrent activity and only training the (far simpler) output weights.

In addition to simplifying the training process, echo state networks shift the problem into a higher dimensional space~\cite{Jaeger2007b}. This technique of dimensional expansion is commonly employed in machine learning, for instance by Support Vector Machines, Multilayer Perceptrons, and many kernel methods. A decision boundary which is nonlinear in the original problem space is often linear in higher dimensions, allowing a more efficient learning procedure to be used~\cite{Campbell2002,hastie2005elements}.

The echo state networks we used here consists of 10 input nodes, 1 output node and  a ``reservoir,'' consisting of 128 hidden nodes, which is randomly and recurrently connected. The connection weights $\mathbf{W}$ within the reservoir as well as the weights to it from the input and output nodes ($\mathbf{W}_\text{in}$ and $\mathbf{W}_\text{fb}$, respectively) are sampled uniformly at random from the interval $[0, 1]$. $\mathbf{W}$ is also scaled such that the spectral radius $\rho(\mathbf{W})<1$~\cite{Buehner2006}. This scaling ensures the network will exhibit the ``echo state property:'' the effect of previous reservoir states and inputs will asymptotically approach zero as time passes rather than persisting indefinitely or being amplified~\cite{Lukovsevivcius2009}. Only the weights $\mathbf{W}_\text{out}$ from the reservoir to the output nodes are trained. The goal is to draw on the diverse set of behaviors within the reservoir and find some linear combination of those oscillations which match the desired output.

States of reservoir nodes $\mathbf{y}_t$ are updated according to
\begin{align}
	\mathbf{y}_t = \sigma( \mathbf{W}_\text{in} \mathbf{x}_t + \mathbf{W} \mathbf{y}_{t-1} + \mathbf{W}_\text{fb} z_{t-1} )
\end{align}
where $\mathbf{x}_t$ is the current network input, $z_{t-1}$ is the previous network output, and $\sigma$ is the logistic sigmoid function. 
The output of the network is determined by
\begin{align}
	z_t = \sigma( \mathbf{W}_\text{out} \left[ \mathbf{x}_t | \mathbf{y}_t \right] )
\end{align}
where $|$ represents a vertical concatenation.

The training procedure involves presenting the network with each input in the sequence and updating the internal reservoir. The inputs and reservoir states are collected row-wise in a matrix $\mathbf{S}$. We redefine the network's targets during training to be $z'_t = \sigma^{-1}(z_t)$ and collect them row-wise in $\mathbf{D}$. This allows us to use a standard pseudo-inverse solution to compute the output weights $\mathbf{W}_\text{out} = (\mathbf{S}^{-1}\mathbf{D})^\mathsf{T}$ which minimizes the MSE of the network on the training output. 

\section{Data Collection and Preprocessing}

The data consists of the Twitter statuses of 12,043 users over a 49 day period. The users are embedded in a 15,000 node network collected by performing a breadth-first expansion from a seed user. Once the seed user was chosen, the network was expanded to include his/her followers, only including users considered to be active (users who tweeted at least once per day over the past one hundred tweets). Network collection continued in this fashion by considering the active followers of the active followers of the seed, etc.

The statuses of each user were transformed into a binary time series using their time stamp, as described in the Methodology section. In this paper, only tweets made between 7 AM and 10 PM (EST) were considered. Since most of the users in our dataset reside on the East Coast of the United States, this windowing was chosen because of the conditional stationarity assumption on $\{ X_{i}\}_{i = 1}^{T}$: users would have different conditional distributions during waking and sleeping hours. For any second during this time window, a user either tweets, or does not. Thus, each day can be considered as a binary time series of length 57,600, with a 1 at a timepoint if the user tweets, and a 0 otherwise.

Because of statistical and computational limitations, the time series were further coarsened by binning together disjoint intervals of time. We considered time windows with length equal to ten minutes ($\Delta t = 600$). Thus, we created a new time series by recording a 1 if any tweeting occurs during a ten minute window, and a 0 otherwise. In theory, this coarsening weakens our predictive ability: in the limit of infinite data, the data processing inequality tells it is always better for prediction to have raw data rather than a function of the data \cite{cover2012elements}. However, because of the practical constraints of finite data and finite computing power, the coarsening of the data allows for the inference of tractable models which are useful in practice. Once we have the (either coarsened or not) time series, we can visualize the behavior of a user over the 49 day period by using a rastergram. A rastergram visualizes a point process over time and over trials. The horizontal axis corresponds to the time point in a particular day, and the vertical axis corresponds to the day number. At each time point, a vertical bar is either present (if the user tweeted on that day at that time) or absent. Visual inspection of rastergrams serves as a first step towards understanding the behavior of any given user. \hyperlink{link:coarsening}{Figure~\ref{Fig-coarsening}} demonstrates the original and coarsened time
\begin{Figure}
\centering
\hypertarget{link:coarsening}{}
\includegraphics[width=\columnwidth]{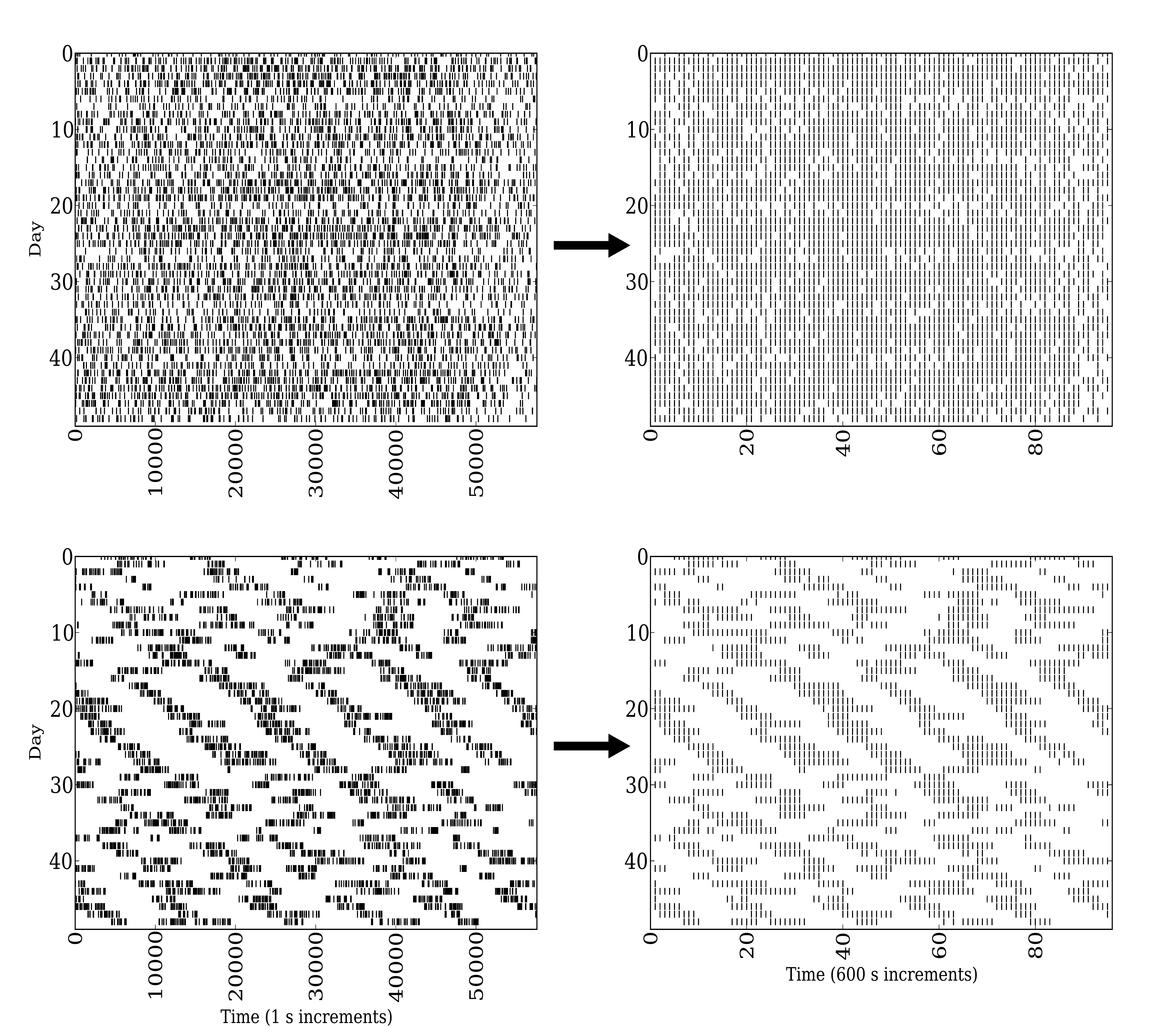}
\captionof{figure}{Coarsening of two users. Each row in the rastergram corresponds to a single day of activity for a fixed user. The original time series are at single second resolution, resulting in 57,600 time points in each day. After binning together activity using disjoint (partitioned) ten minute windows, there are 96 time points in each day ($T = 96$).}
\label{Fig-coarsening}
\end{Figure}
series for two users.
 
The users were further filtered to include only the top 3,000 most active users over the 49 day period. A base activity measure was determined by the proportion of seconds in the 7 AM to 10 PM window the user tweeted, which we call the tweet rate. Of the top 3,000 users, these tweet rates ranged from $0.38$ to $8.5 \times 10^{-5}.$ 90\% of the top 3,000 users had a tweet rate below 0.05. The distribution of the tweet rates amongst the top 3,000 users is shown in \hyperlink{link:hist_tweetrate}{Figure~\ref{Fig-hist_tweetrate}}.

\section{Results and Discussion}

\subsection{Testing Procedure}

The 49 days of user activity were partitioned, chronologically, into a 45 day training set and a 4 day testing set. This partition was chosen to account for possible changes in user behavior over time, which would not be captured by using a shuffling of the days. Thus, for each user, the training set consists of 4,320 timepoints, and the testing set consists of 384 timepoints.

The only parameter for the causal state model is the history length $L$ to use. This was treated as a tuning parameter, and the optimal value to use was determined by using 9-fold cross-validation on the training
\begin{Figure}
\centering
\hypertarget{link:hist_tweetrate}{}
\includegraphics[width=\columnwidth]{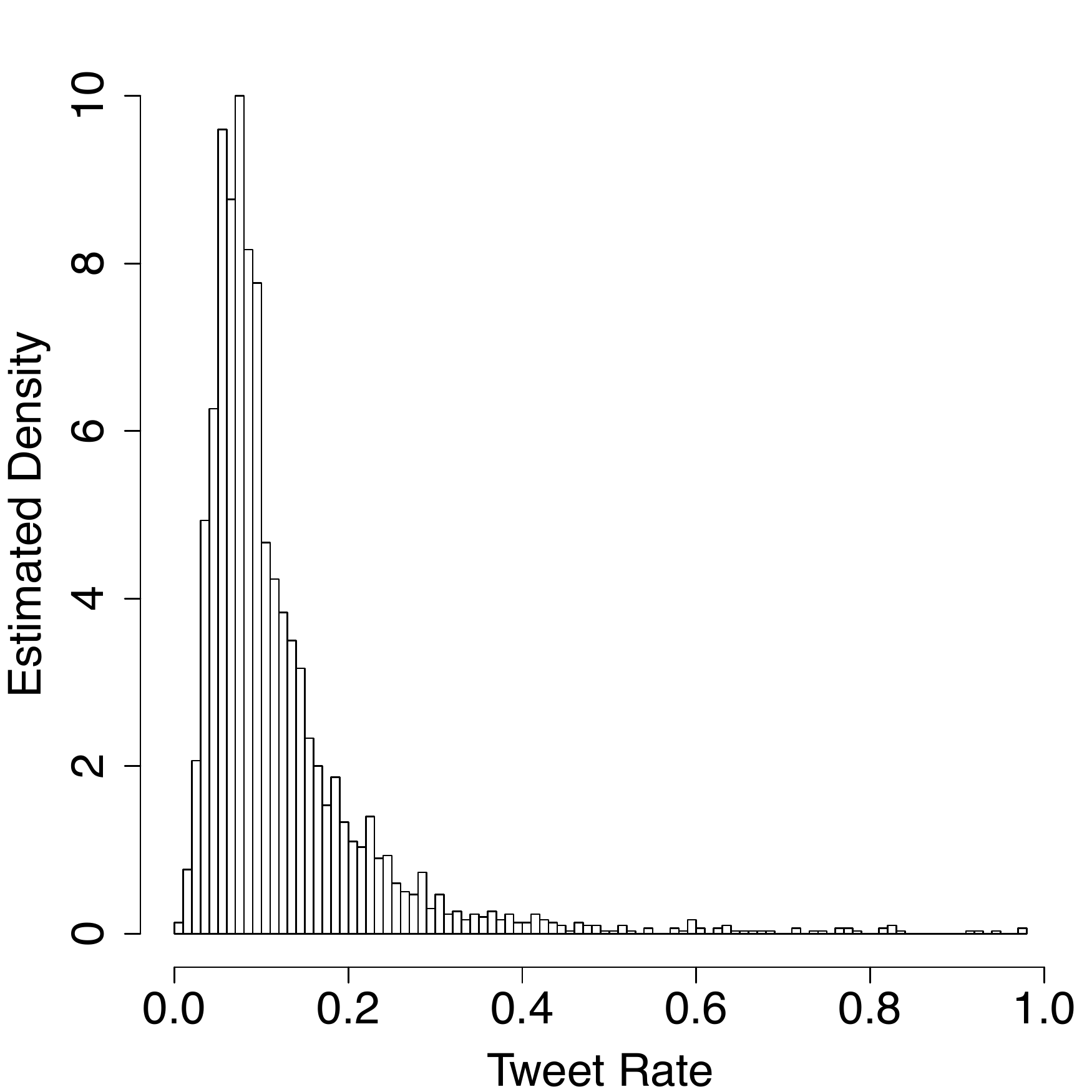}
\captionof{figure}{The observed distribution of the fraction of time spent tweeting (tweet rate) over the 49 day period for all of the users. 90\% of the 3,000 users had a tweet rate below 0.05.}
\label{Fig-hist_tweetrate}
\end{Figure}
 set. The maximal history length $L_{\text{max}}$ that can be used and still ensure consistent estimation of the joint distribution is dependent on the number of time points $n$, and is bounded by
\begin{align}
	L_{\text{max}} < \frac{\log_{2} n}{h + \epsilon},
\end{align}
where $h$ is the entropy rate of the stochastic process and $\epsilon$ is some positive constant~\cite{marton1994entropy}. Thus, because $0 \leq h \leq 1$ for a stationary stochastic process with two symbols, as a practical bound, we take 
\begin{align*}
	L_{\text{max}} < \log_{2} n.
\end{align*}
For this data set, the bound requires that $L_{\text{max}} < 12$. Thus, we use the 9-fold cross-validation to reconstruct causal state models using histories of length 0 through 11, and then choose the history length that maximizes the average accuracy rate taken over all of the folds.

Experiments showed that the echo state network was robust to varying parameter choices as long as the echo state property is achieved~\cite{Ozturk2007,Rodan2011}. As a result all networks were created with $\rho(\mathbf{W}) = 0.99$ and $L_{\text{ESN}} = 10$.

\subsection{Comparison to Baseline}

In all cases, we compute the accuracy rate of a predictor using zero-one loss. That is, for a given user, we predict the time series $X_{1}, \ldots, X_{n_{\text{test}}}$ as $\hat{X}_{1}, \ldots, \hat{X}_{n_{\text{test}}}$ and then compute
\begin{align}
	\text{Accuracy Rate} = \frac{1}{n_{\text{test}}} \sum_{i = 1}^{n_{\text{test}}} 1 [\hat{X}_{i} = X_{i}].
\end{align}

We compared the accuracy rates on the causal state model and echo state network to a baseline accuracy rate for each user. The baseline predictor was taken to be the majority vote of tweet vs.\@ not-tweet behavior over the training days, regardless of the user's past behavior. That is, for the baseline predictor we take
\begin{align}
	\hat{X}_{i} = \left \{\begin{array}{cc}
		0 &:  \hat{p} \leq \frac{1}{2} \vspace{0.2cm}\\
		1 &:  \hat{p} > \frac{1}{2}
	\end{array} \right.,
\end{align}
where 
$\hat{p} = \frac{1}{n_{\text{train}}} \sum_{j = 1}^{n_{\text{train}}} X_{j}.$
This is the optimal predictor for a Bernoulli process where the $\{ X_{i}\}$ are independent and identically distributed Bernoulli random variables with parameter $p$. In the context of our data, for users that usually tweeted in the training set, the baseline predictor will always predict that the user tweets, and for users that usually did not tweet in the training set, the baseline predictor will always predict the user does not tweet. For any process with memory, as we would expect from most Twitter users, a predictor should be able to outperform this base rate.

The comparison between the baseline predictor and the causal state model and echo state network predictors are shown in \hyperlink{link:Improvement}{Figure~\ref{Fig-Improvement}}. In both plots, each red point corresponds to the baseline rate on the testing set for a given user, and the blue point corresponds to the accuracy rate on the testing set using one of the two models. Here, the tweet rate is computed in terms of the coarsened time series. That is, the tweet rate is the proportion of ten minute windows over the 49 day period which contain one or more tweet. Clearly, the model predictions show improvement over the baseline prediction, especially for those users with a tweet rate above 0.2.

To make this more clear, the improvement as a function of the tweet rate of each user is shown in \hyperlink{link:tweetrate_v_improvement}{Figure~\ref{Fig-tweetrate_v_improvement}} for both methods. Breaking the users into two groups, with the high tweet rate group having a tweet rate greater than 0.2 and the low tweet rate group having a tweet rate greater than or equal to 0.2, we can estimate the conditional density of improvements among these groups. These estimated densities are shown in \hyperlink{link:kde_improvement_by_tweetrate}{Figure~\ref{Fig-kde_improvement_by_tweetrate}}. We see that most of the improvement lies in the high tweet rate group, while the low tweet rate group is concentrated around 0 improvement.

\begin{Figure}
\vspace{-2.5mm}
\hypertarget{link:Improvement}{}
\includegraphics[width=\columnwidth]{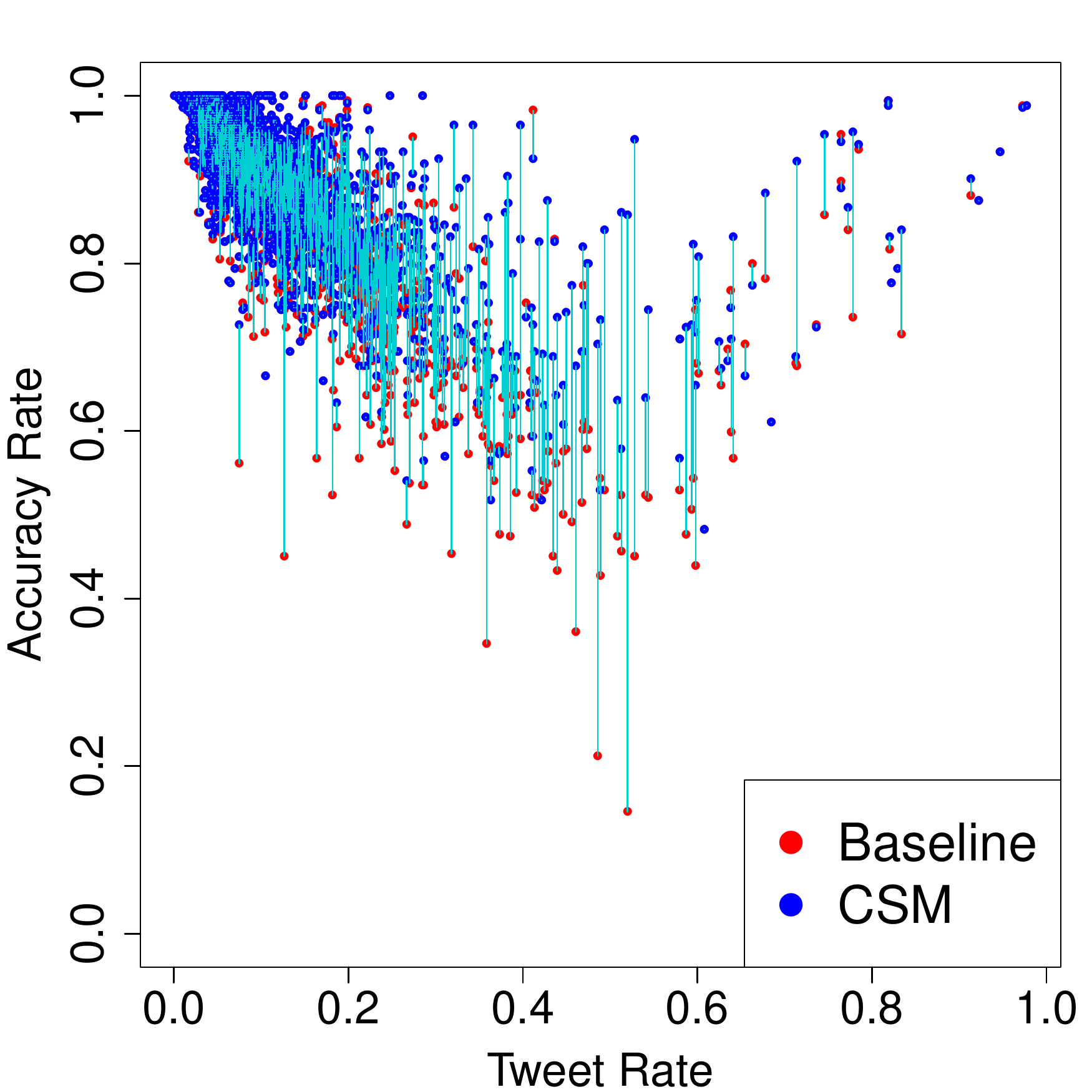}
\includegraphics[width=\columnwidth]{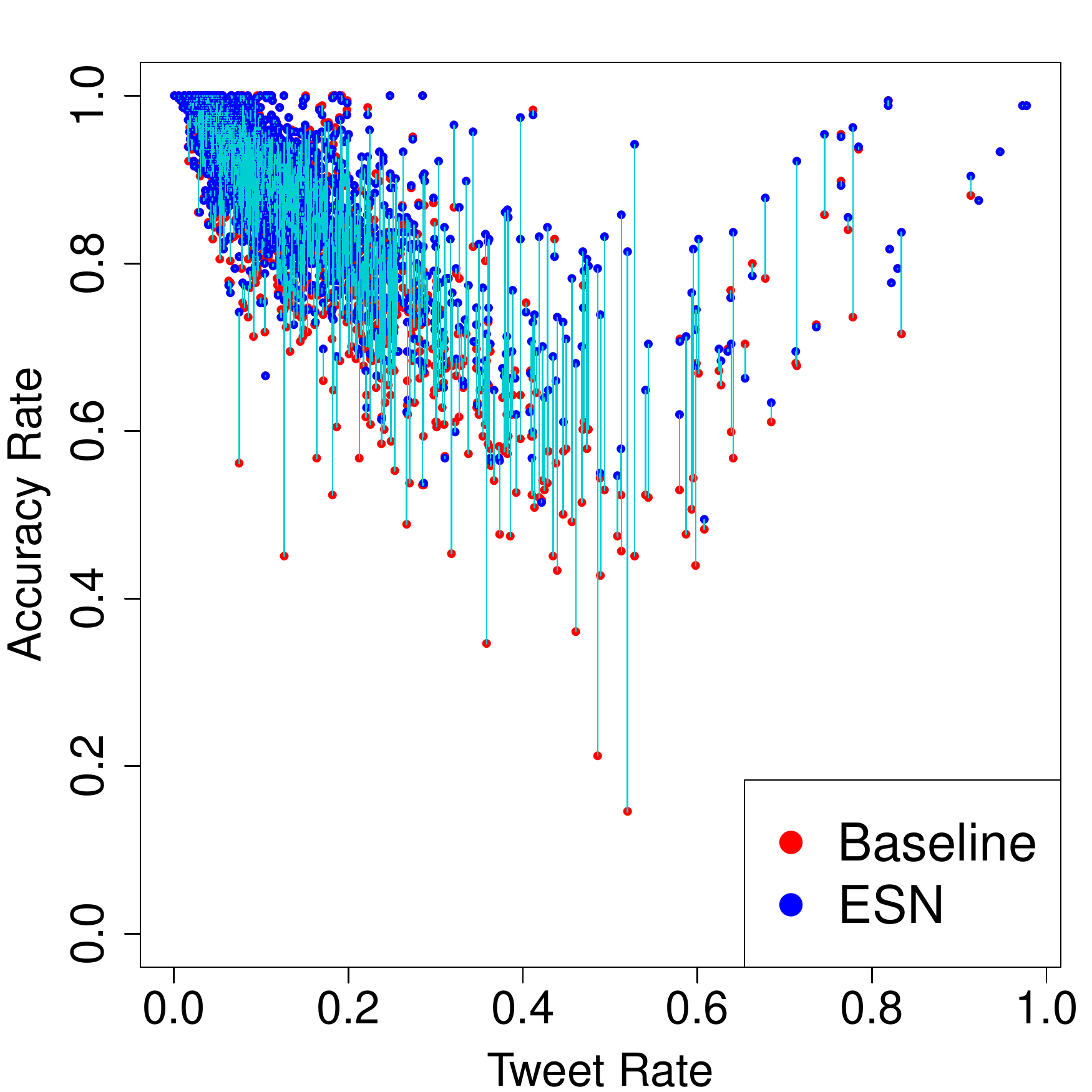}
\captionof{figure}{The improvement over the baseline accuracy rate for the causal state model (top) and echo state network (bottom). In both plots, each red point corresponds to the baseline accuracy rate for a user, and the connected blue point is the accuracy rate using either the causal state model or the echo state network.}
\label{Fig-Improvement}
\end{Figure}

\subsection{Typical Causal State Models for the Users}

The causal states $\{S_{i}\}_{i = 1}^{T}$ of a stochastic process $\{X_{i}\}_{i = 1}^{T}$ form a Markov chain, and the current causal state $S_{i}$ plus the next emission symbol $X_{i + 1}$ completely determine the next causal state $S_{i + 1}$~\cite{shalizi2001computational}.
\begin{Figure}
\hypertarget{link:tweetrate_v_improvement}{}
\includegraphics[width=\columnwidth]{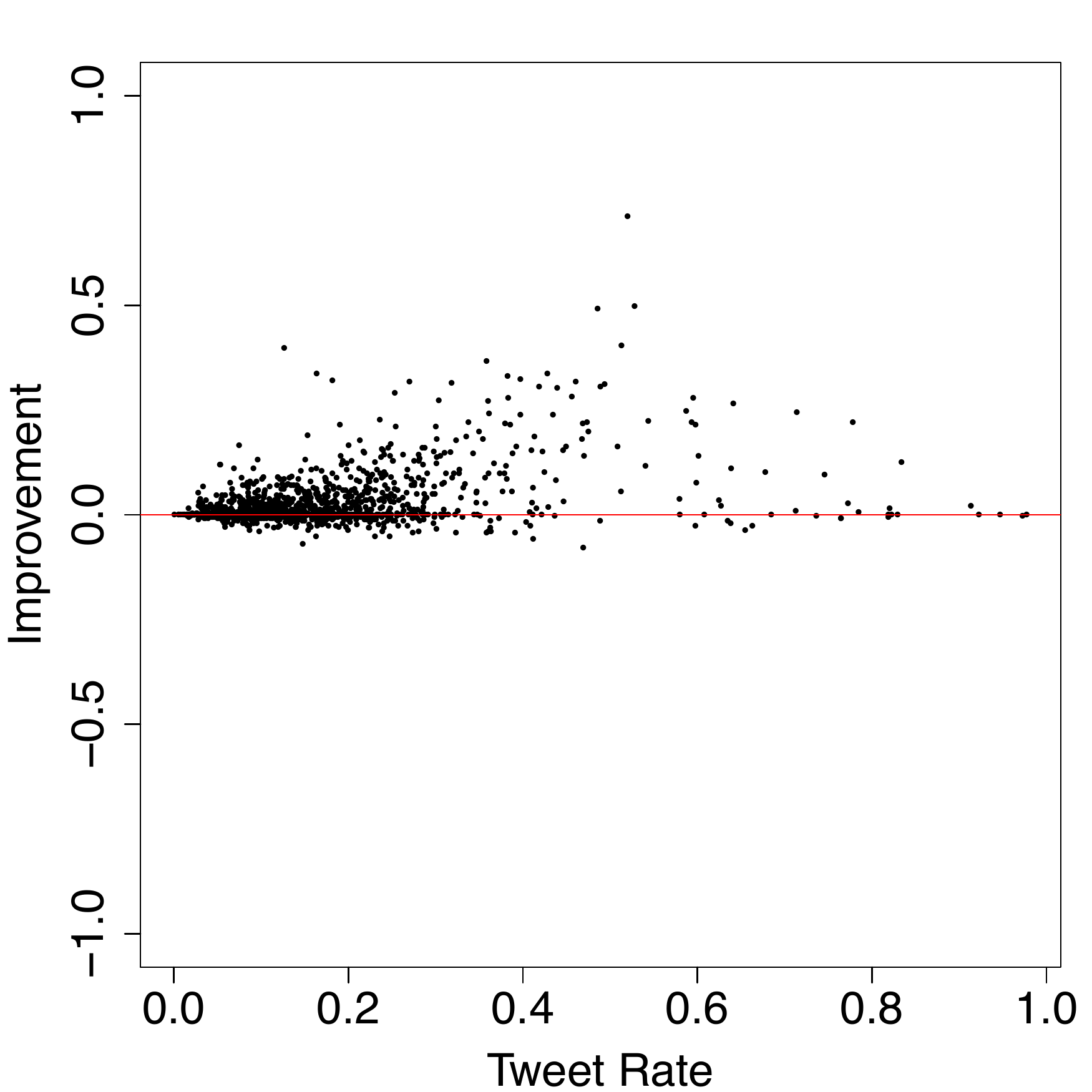}
\includegraphics[width=\columnwidth]{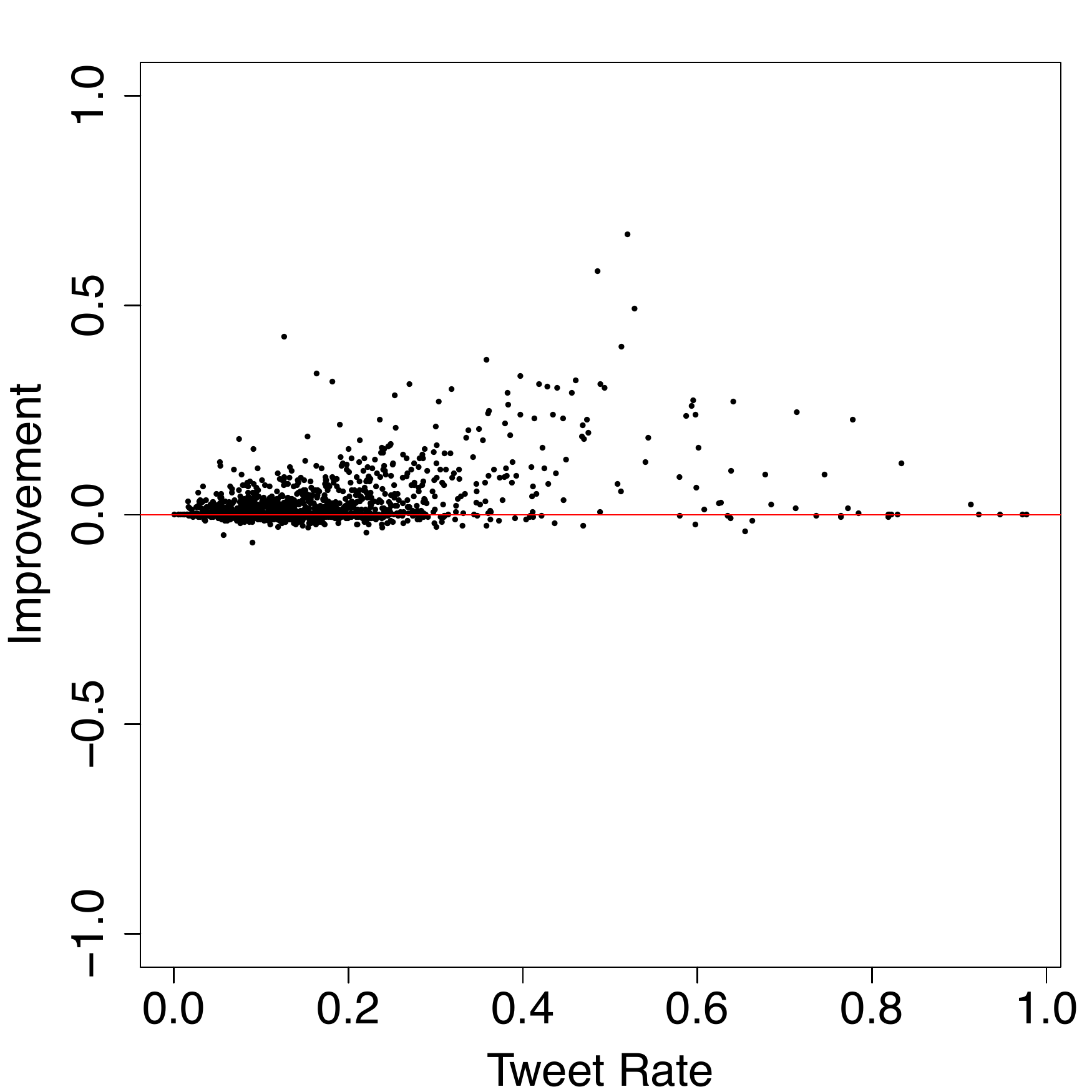}
\captionof{figure}{The improvement over the baseline accuracy rate for the causal state model (top) and the echo state network (bottom). For both models, the greatest improvement occurred for a coarsened tweet rate near $\frac{1}{2}.$}
\label{Fig-tweetrate_v_improvement}
\end{Figure}
 These two properties of a causal state model allow us to write down an emission-decorated state-space diagram for a given user. That is, the diagram resembles the state-space diagram for a Markov (or Hidden Markov) model, with the additional property that we must decorate each transition between states by the symbol emitted during that transition.
 
Several such diagrams are shown in \hyperlink{link:sample_CSMs}{Figure~\ref{Fig-sample_CSMs}}. Each circle corresponds to a causal state, and each arrow corresponds to an allowable transition. The arrows
\begin{Figure}
\hypertarget{link:kde_improvement_by_tweetrate}{}
\includegraphics[width=\columnwidth]{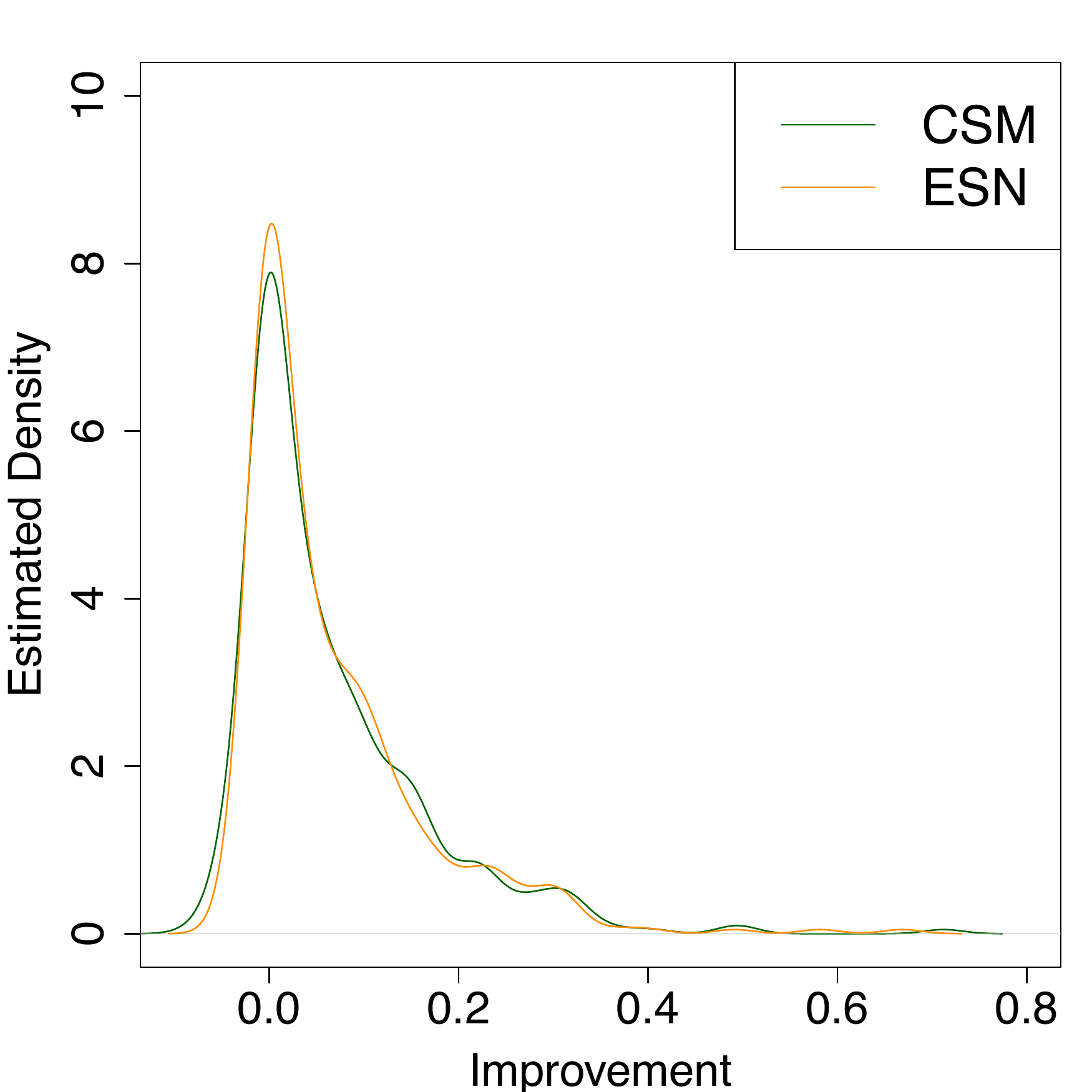}
\includegraphics[width=\columnwidth]{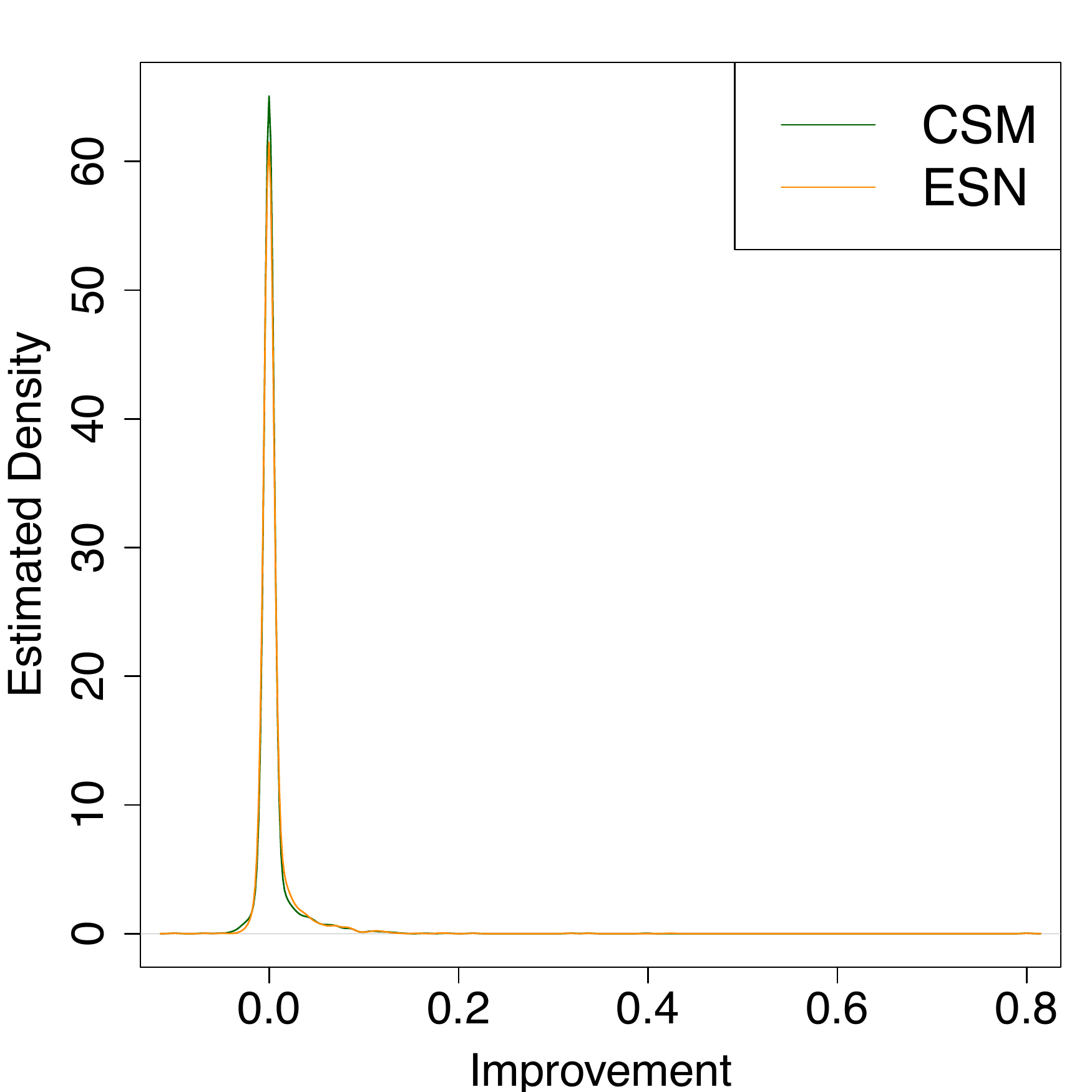}
\captionof{figure}{The distribution of improvements for both the causal state model (top) and echo state network (bottom), with the users partitioned into `High Tweet Rate' (tweet rate greater than 0.2) and `Low Tweet Rate' (tweet rate lower than 0.2) groups.}
\label{Fig-kde_improvement_by_tweetrate}
\end{Figure}
are decorated with $e_{ij} | p_{ij}$, where $e_{ij}$ is the emission symbol observed transitioning from causal state $i$ to causal state $j$, and $p_{ij}$ is the probability of transitioning from causal state $i$ to causal state $j$. For example, \hyperlink{link:sample_CSMs}{Figure~\ref{Fig-sample_CSMs}(a)} corresponds to a Bernoulli random process with success probability $p$. At each time step, the causal state returns to itself, emitting either a 1, with probability $p$, or a 0, with probability $1 - p$.

\begin{Figure}
\hypertarget{link:sample_CSMs}{}
\centering
\includegraphics[width=\columnwidth]{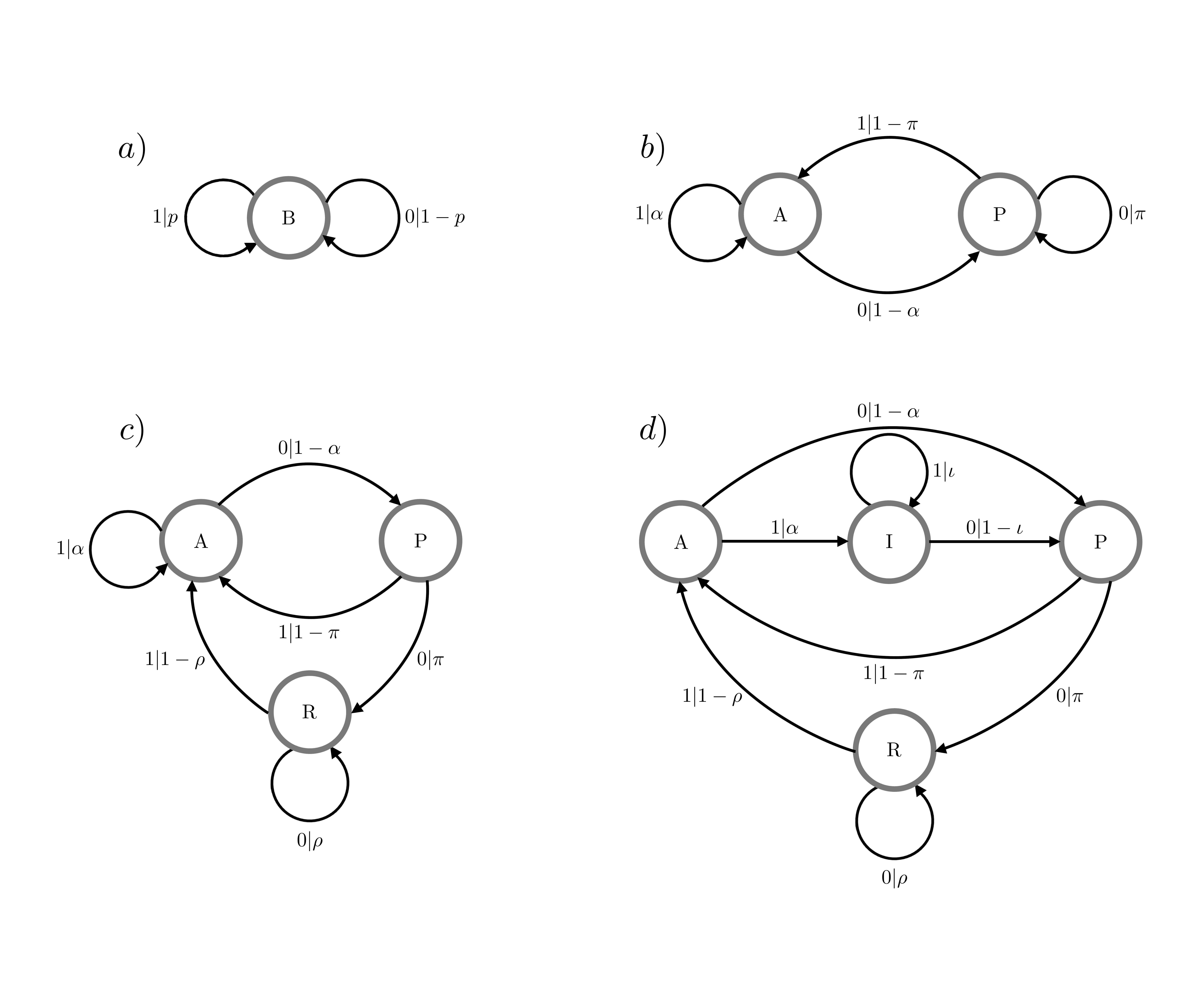}
\captionof{figure}{Typical 1, 2, 3, and 4-state causal state models. Of the 3,000 users, 383 (12.8\%), 1,765 (58.8\%), 132 (4.4\%), and 100 (3.3\%) had these number of states, respectively.}
\label{Fig-sample_CSMs}
\end{Figure}

The four causal state models shown are typical examples of the models observed in 79.3\% of the 3,000 users. The model corresponding to \hyperlink{link:sample_CSMs}{Figure~\ref{Fig-sample_CSMs}(a)} is simple: the user shows no discernible memory and so the behavior is a biased coin flip. Only 383 (12.8\%) of the users correspond to this model. The second model, \hyperlink{link:sample_CSMs}{Figure~\ref{Fig-sample_CSMs}(b)}, displays more interesting behavior. We see that such users have two states, labeled A (active) and P (passive). While the user is in state A, it may stay in state A, continuing to emit 1s, or transition to state P emitting a 0. While in state P, the user may stay in state P, continuing to emit 0s, or transition to state A emitting a 1. Thus, these two states correspond to a user that is typically active or passive over periods of time, exhibiting `bursting' behavior as in the second user in \hyperlink{link:coarsening}{Figure~\ref{Fig-coarsening}}.

Users corresponding to the causal state models shown in \hyperlink{link:sample_CSMs}{Figure~\ref{Fig-sample_CSMs}(c)} and \hyperlink{link:sample_CSMs}{Figure~\ref{Fig-sample_CSMs}(d)} exhibit even more involved behavior. Both have a rest state R, where the user does not tweet. However, the active states show more structure. For example, in \hyperlink{link:sample_CSMs}{Figure~\ref{Fig-sample_CSMs}(c)} we see that the user has an active state A, but sometimes transitions to state P emitting a 0, where the user can then return back to the active state A or transition to the rest state R. \hyperlink{link:sample_CSMs}{Figure~\ref{Fig-sample_CSMs}(d)} shows similar behavior, but with an additional intermediate state I. While these models match our intuitions about how a typical Twitter user might behave, it is important to note that the models result entirely from applying CSSR to the data, and did not require any \emph{a priori} assumptions beyond conditional stationarity.

\subsection{Direct Comparison between the Performance of the Causal State Models and the Echo State Networks}

Given the striking similarity in performance between the causal state model and the echo state network, we next compared them head-to-head on each user. The improvement for the causal state model vs.\@ the improvement for the echo state network on each user is shown in \hyperlink{link:improvement_comparison_scatter}{Figure~\ref{Fig-improvement_comparison_scatter}}. As expected given the previous results, the improvements for each method are very strongly correlated.

Next, we investigated the top 20 users for which the causal state model or the echo state network outperformed the other model. For those users where the causal state model outperformed, the clearest indicator was the structured (near deterministic) behavior of the users. The top four such users are shown in \hyperlink{link:rasters_best-csm}{Figure~\ref{Fig-rasters_best-csm}}. The causal state model inferred from the data can be used to characterize the structure of the observed dynamics in a formal manner~\cite{shalizi2001computational}. Because the hidden states $\mathcal{S} = \{s_{1}, \ldots, s_{|\mathcal{S}|} \}$ determine the observed dynamics, the entropy over those states can be used to characterize the diversity of behaviors a process is capable of. The entropy over the causal state process is called the \emph{statistical complexity} of the process, and given by
\begin{align}
	C &= H[S] \\
	&= - \sum_{s \in \mathcal{S}} P(S = s) \log_{2} P(S = s).
\end{align}
Informally, it is the number of bits of the past of a process necessary to optimally predict its future. For example, for an IID process, $C = 0$, since none of the past is necessary to predict the future, while for a period-$p$ process, $C = \log_{2} p$, since it takes $\log_{2} p$ bits of the past to synchronize to the process.

Of the top twenty users best predicted by the causal state model, the average statistical complexity was 3.99, while the top twenty users best predicted by the echo state network had an average statistical complexity of 2.72. \hyperlink{link:Cmu_v_comparison}{Figure~\ref{Fig-Cmu_v_comparison}} shows the difference between the two methods as a function of the inferred statistical complexity. We see that the causal state models tend to outperform the echo state network for high statistical complexity users, while the echo state network tends to outperform for the low (near 0) statistical complexity users.

Of the top twenty users best predicted by the echo state network, we observed that the test set tended to
\begin{Figure}
\hypertarget{link:improvement_comparison_scatter}{}
\centering
\includegraphics[width=\columnwidth]{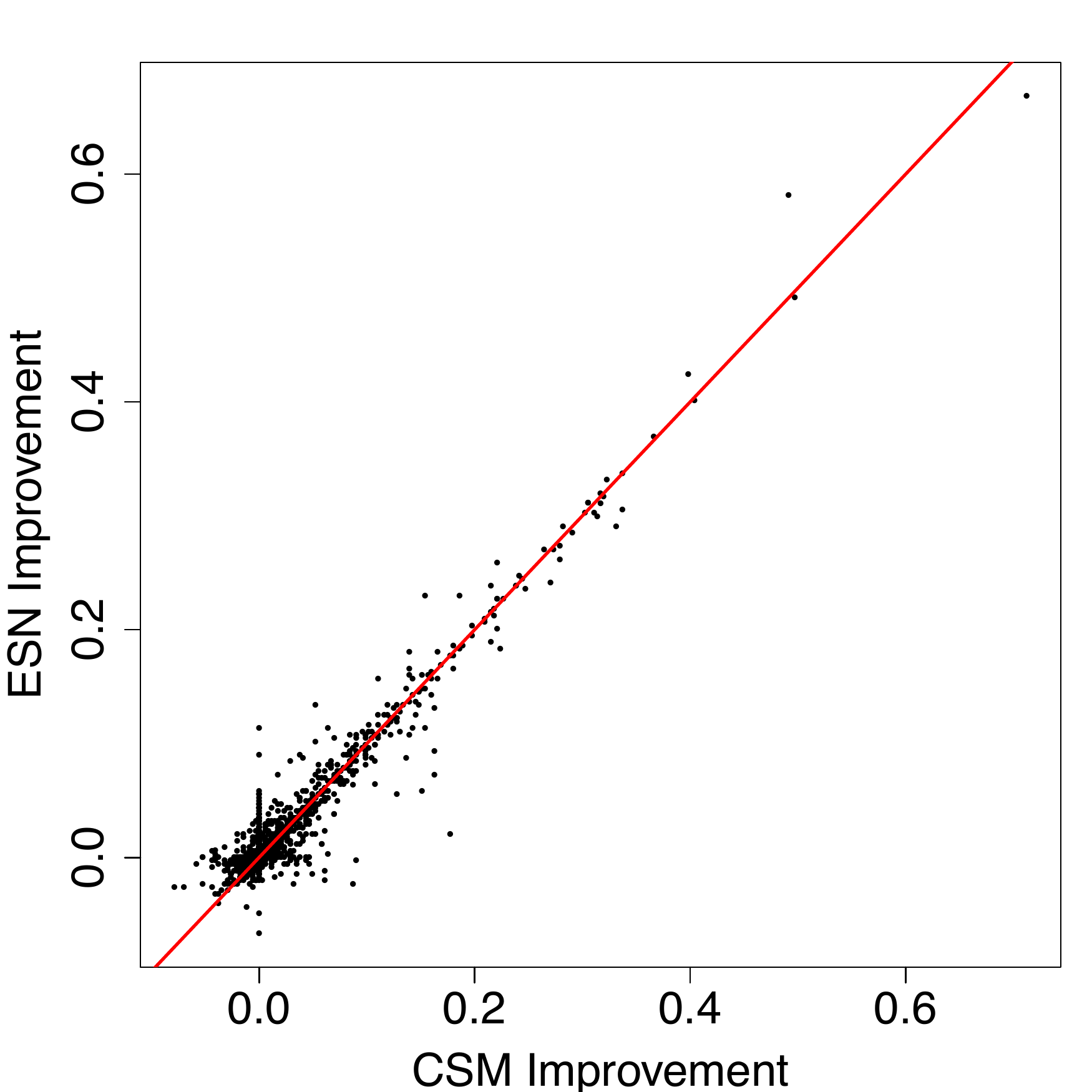}
\captionof{figure}{The improvement over baseline for the causal state model vs.\@ the improvement over baseline for the echo state network. The red line indicates identity, where the two methods improve equally over the baseline predictor.}
\label{Fig-improvement_comparison_scatter}
\end{Figure}
\begin{Figure}
\hypertarget{link:rasters_best-csm}{}
\centering
\includegraphics[width=\columnwidth]{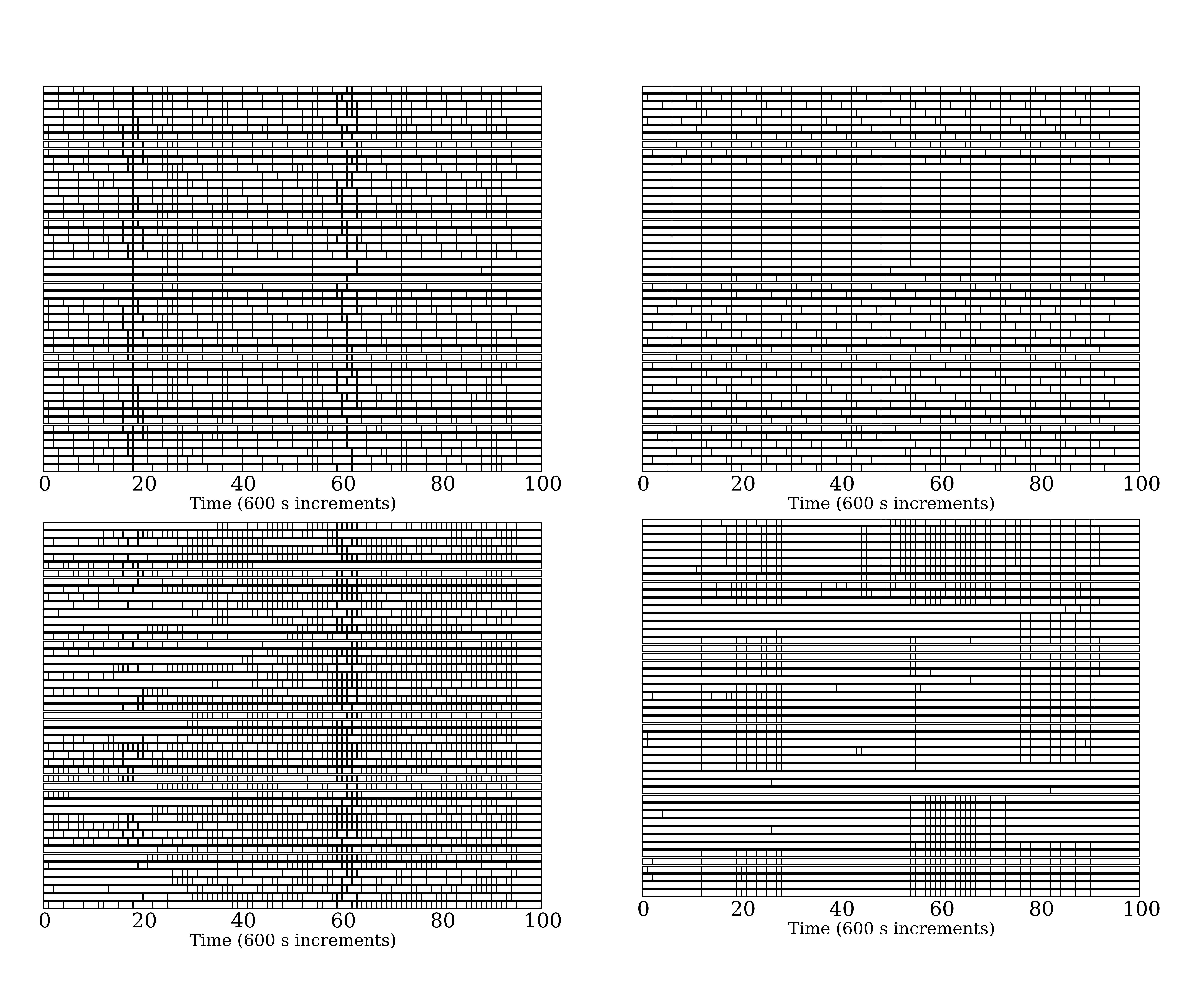}
\captionof{figure}{Raster plots for the four users where the causal state model most outperformed the echo state network. Note that in all but the bottom left case, the users show highly `patterned' behavior. This is typical of the top twenty users for which the causal state model outperformed the echo state network.}
\label{Fig-rasters_best-csm}
\end{Figure}
differ from the training set. To test this hypothesis, we estimated the entropy rates of the test and training sets. The entropy rate $h$ of a stochastic process $\{ X_{i}\}_{i = 1}^{\infty}$ is defined as the limit of the block entropies of length $L$ as the block length goes to infinity,
\begin{align}
	h = \lim_{L \to \infty} \frac{1}{L} H[X_{1}, \ldots, X_{L}].
\end{align}
Thus, the entropy rate can be approximated by estimating block entropies
\begin{align}
	H_{L} = \frac{1}{L} H[X_{1}, \ldots, X_{L}]
\end{align}
of larger and larger block sizes and observing where the block entropies asymptote, as they must for a stationary stochastic process~\cite{cover2012elements}. Unlike block-1 entropy (Shannon entropy), the entropy rate accounts for long range correlations in the process that may explain apparent randomness.

As we observed in the top twenty users, we see that overall the causal state model tends to perform best relative to the echo state network when the training and test set are similar, while the echo state network tends to outperform in the cases where the training and test set differ. This can be seen in \hyperlink{link:abs_diff_entropy_quartile}{Figure~\ref{Fig-abs_diff_entropy_quartile}}, in which the users have been grouped into quartiles by the absolute value of the difference between training and test set entropy rates.

\subsection{Bit Flip Experiment}

To further explore this difference between the two models, we performed the following `bit flip' experiment. For each user, we trained both the causal state model and the echo state network on the full 49 days of data. We then tested the users on the same data, but with some proportion $q$ of data set flipped such that 0s become 1s and vice versa, with $q$ ranging from 0 to 1 in increments of 0.1. This allows us to synthetically create examples where the training and test sets differ as much or as little as desired by systematically adding noise into the time series.

The result of this experiment is shown in \hyperlink{link:bitflip_all}{Figure~\ref{Fig-bitflip_all}}. The causal state model performs as expected, with the accuracy rate degrading as the corruption in the training set approaches 50\%. Beyond this point, the large variance in the accuracy rates result from the different types of models inferred from the data. In particular, the 58.8\% of users with a two-state `bursting' causal state model as in \hyperlink{link:sample_CSMs}{Figure~\ref{Fig-sample_CSMs}(b)} continue to perform well, as the recoding of a burst of zeros or ones does not effect the predictive capability of the model. 

The echo state networks show the same degradation in accuracy rate as the corruption in the training set
\begin{Figure}
\hypertarget{link:Cmu_v_comparison}{}
\centering
\includegraphics[width=\columnwidth]{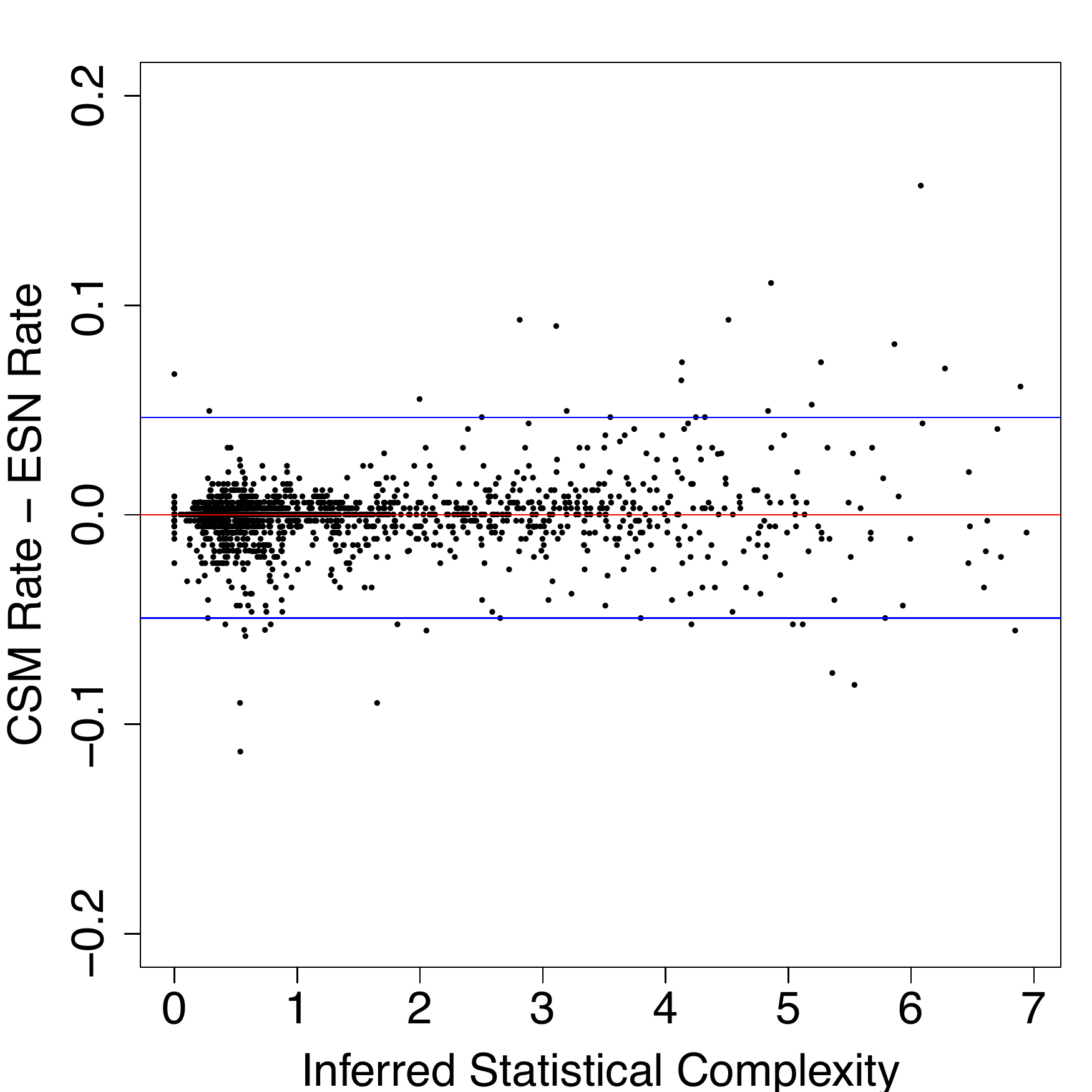}
\captionof{figure}{The difference in improvement between the causal state model and the echo state network for each user as a function of the inferred statistical complexity $C$ of each user. The blue lines indicate the cutoff points above and below which the top twenty best users for the causal state model and echo state network, respectively, lie, and correspond to 0.0465 and -0.0494.}
\label{Fig-Cmu_v_comparison}
\end{Figure}
\begin{Figure}
\centering
\vspace{-5mm}
\hypertarget{link:abs_diff_entropy_quartile}{}
\includegraphics[width=\columnwidth]{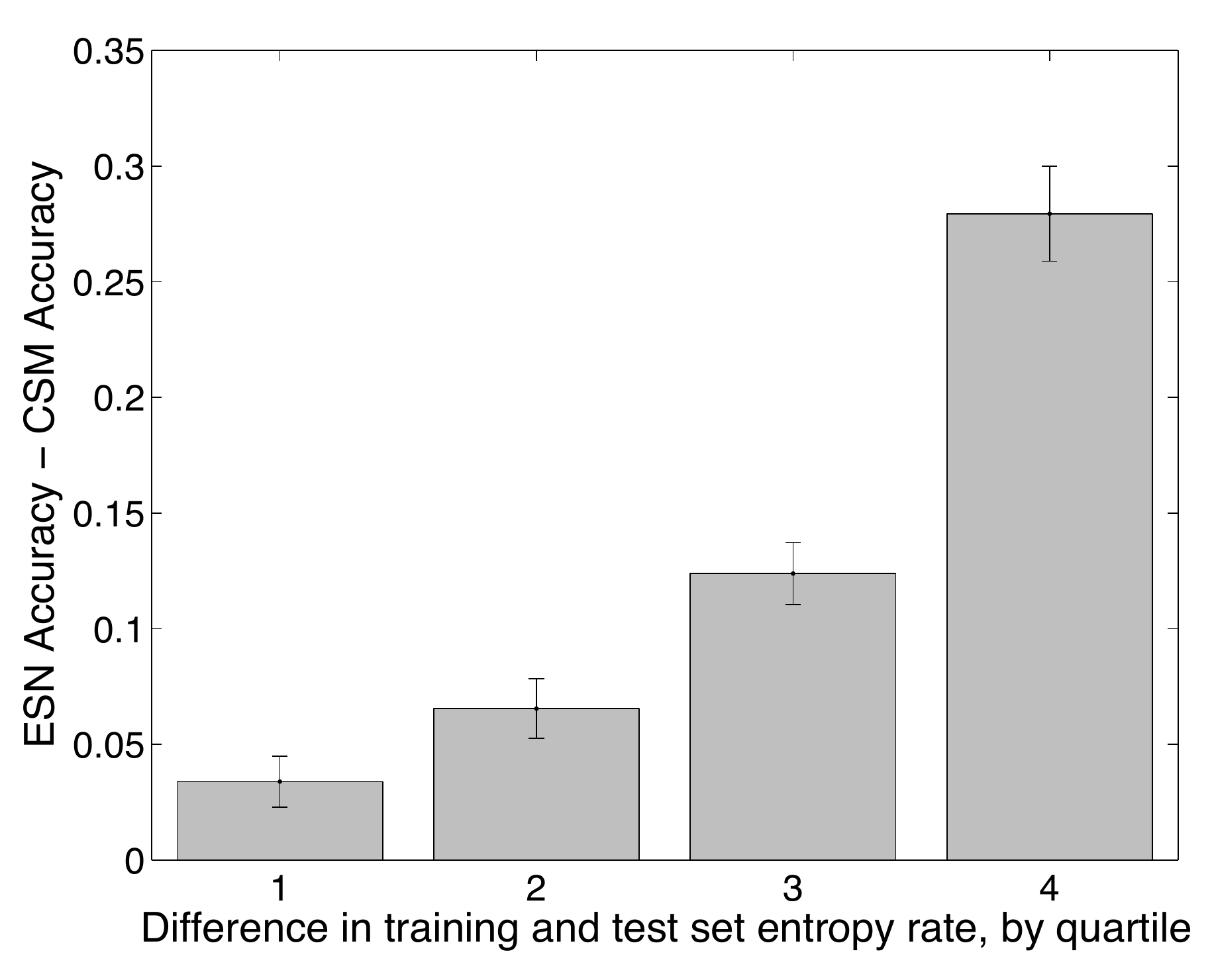}
\captionof{figure}{The difference in accuracy rates between the causal state model and the echo state network for each user, binned into quartiles by the absolute value of the difference in entropy rates for the training and testing sets. The causal state model performs best when this difference is low, and the echo state network performs best when it is high.
}
\label{Fig-abs_diff_entropy_quartile}
\end{Figure}
approaches 50\%, but beyond this amount they begin to show improvement. The large variance in the accuracy rates is again explained by a bimodality in the accuracy rates.

\begin{Figure}
\hypertarget{link:bitflip_all}{}
\centering
\includegraphics[width=\columnwidth]{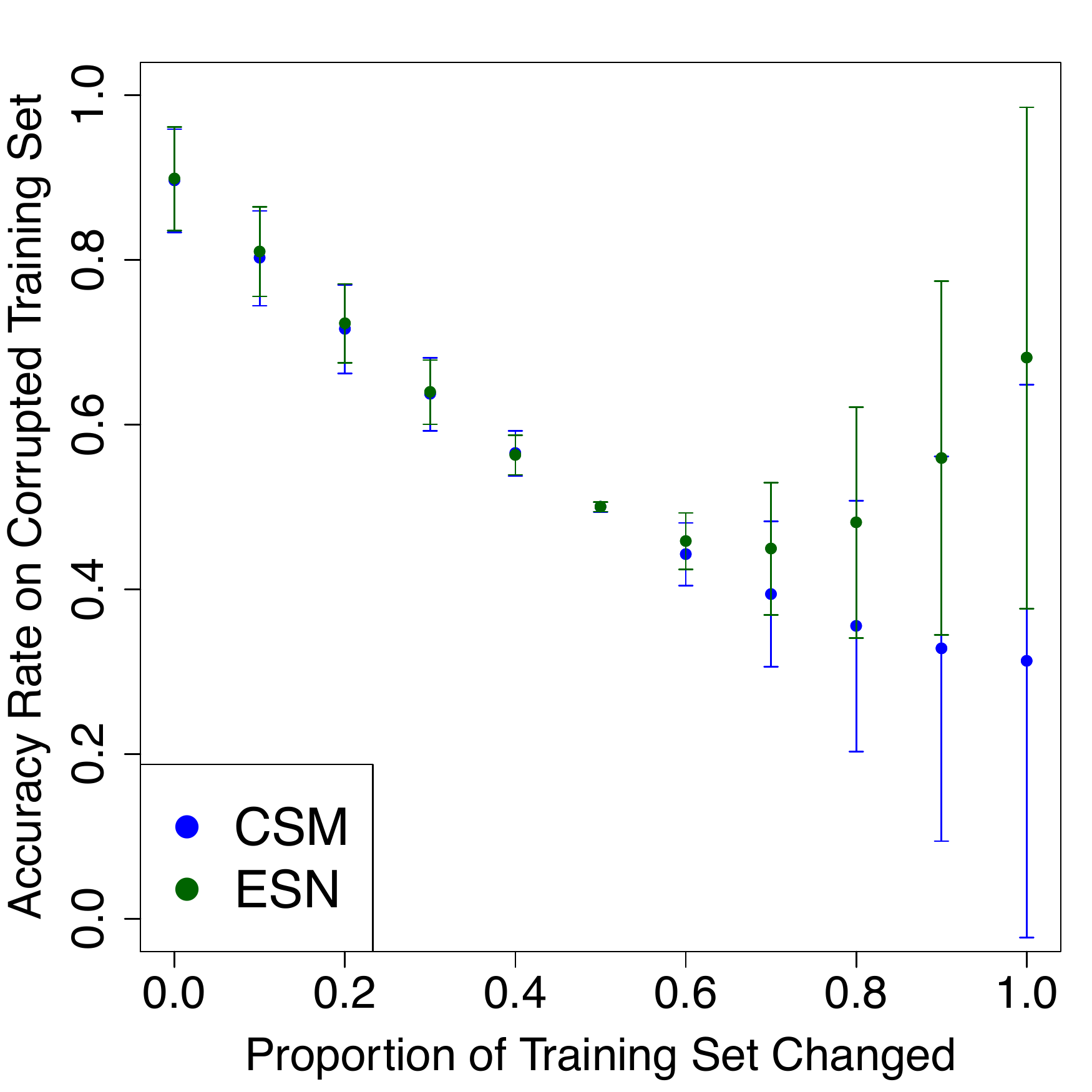}
\captionof{figure}{The accuracy rate of the causal state model and echo state network tested on its training data, with the training data corrupted by flipping a proportion $q$ of the bits. Bars indicate plus or minus one standard deviation in the accuracy rates across all users.}
\label{Fig-bitflip_all}
\end{Figure}

We believe the improvement in accuracy the echo state networks display when more than 50\% of bits are changed is a result of many of the networks having learned a simple ``trend-following'' model: if you are in a tweeting state, continue tweeting; if you are in a non-tweeting state, continue not tweeting. This is very similar to the commonly observed two-state causal state model (\hyperlink{link:sample_CSMs}{Figure~\ref{Fig-sample_CSMs}(b)}) with one important difference\,---\,the echo state network does not fix the probabilities of being in either the active or passive states based on the training data. When a high proportion of bits have been flipped a sequence of, for instance, short periods of activity embedded in long stretches of quiescence will become the inverse: short periods of silence and long stretches of activity. A causal state model which has learned a two-state solution based on the original data will struggle since it expects different probabilities than those observed in the corrupted sequences, while an echo state network that has learned only to follow the recent trend will be able to adapt to the new, altered sequences so long as there are long trends remaining in the data. The echo state network thus displays less fidelity to the observed data, but in doing so may be better able to adapt to particular perturbations if the patterns change, for example a user who maintains a `bursting' pattern over time, but changes the length of these bursts.

\subsection{Overview of Results}

Overall, the causal state models and the echo state networks both showed improvement, and in some cases drastic improvement, over a baseline predictor. Moreover, for a large proportion of the users, the two methods gave very similar predictive results, as exemplified by \hyperlink{link:improvement_comparison_scatter}{Figure~\ref{Fig-improvement_comparison_scatter}}. Out of all the users, 58.8\% had inferred causal state models similar to \hyperlink{link:sample_CSMs}{Figure~\ref{Fig-sample_CSMs}(b)}, where a user has a tweeting state A and a non-tweeting state P. This bursting-type behavior is naturally captured by the echo state network, and thus the similarity in performance on these users is to be expected.

We have observed that predictability of user behavior is not homogeneous across the 3,000 users considered, and in many cases the \emph{reason} for the difficulty in prediction differs across users. In some cases, considering a long enough history of a user's behavior is enough to predict their future behavior, but others still appear random after accounting for previous behavior. 

\section{Conclusion and Future Work}







In this paper, we have shown that by building representations of the latent states of user behavior we can start to predict their actions on social media.
We have done this using two different approaches, which have different ways of capturing the complexity of user behavior. Causal state modeling starts from a simple model and adds structure, while echo state networks start with complex descriptions and simplify relationships.
We hypothesized that these two methods would perform differently when applied to a diverse collection of users derived from a real world social media context.
Our results indicate that the two methods perform differently under different conditions. Specifically, computational mechanics provides a better model of a user's behavior when it is highly structured and does not change dramatically over time, while the echo state network approach seems to be more adaptive, while at the same time giving up some of the deep structure present in the behavior.
Moreover, we have shown that both methods are robust to noise and decay gracefully in performance.

Ultimately, the two methods performed very similarly on a large proportion of the users.
It should be noted that this was not expected.
The two methods differ drastically in their modeling paradigm, and the data was quite dynamic, providing plenty of opportunity for differentiation.
Our best explanation is that in the end, and as noted above, most users exhibit only a few latent states of behavioral processing, and as such any model which is able to capture these states will do well at capturing the behavior of users.
We could test this hypothesis in future work by restricting the number of states that both the echo state network and the computational mechanics approach can use, and observing if the results change substantially.

However, before we address that question, there are several other limitations of the present work that need to be addressed.
One of the biggest weaknesses of the present approach is its failure to incorporate exogenous inputs to a user. That is, we have treated each user as an autonomous unit, and only focused on using their own past behavior to predict their future behavior. In a social context, such as Twitter, it makes more sense to incorporate network effects, and then examine how the behavior of friends and friends of friends directly impact a user's behavior. For example, the behavior of many of the users, especially those users with a low tweet rate, may become predictable after incorporating the behavior of users in their following network. The computational mechanics formalism for doing so has been developed in terms of random fields on networks~\cite{shalizi2003optimal} and transducers~\cite{shalizi2001causal}, but it has yet to be applied to social systems.

We have also simplified the problem down to its barest essentials, only considering whether a tweet has occurred and not its content. Information about the content of a tweet should not \emph{decrease} the predictive abilities of our methods, and could be incorporated in future work, for example, by extending the alphabet of symbols which we allow $X_{i}$ to take.

This study has also focused on user behavior over a month and a half period. With additional data, a longitudinal study of users' behaviors over time could be undertaken. We have implicitly assumed the conditional stationarity of behavior in our models, but these assumptions could be tested by constructing models over long, disjoint intervals of time and comparing their structure.

We have seen that taking a predictive, model-based approach to exploring user behavior has allowed us to discover typical user profiles that have predictive power on a popular social media platform.
Moreover, we have shown this using two different modeling paradigms.
In the near future, we plan to extend this work to take into account the social aspects of this problem, and see how network effects influence user behavior. However, the increase in predictive power \emph{without} explicitly incorporating social factors gives us reason to believe that it is possible to make predictions in the context of user interactions in social media.
Such predictions, which take into account social context, could be useful in any number of domains.
For instance, in a marketing type approach, these models could be used to understand who will respond to a message that is sent out to a group of users, and potentially even assist in the determination of whether or not a particular piece of content will go viral.
Predicting user behavior on social media has the potential to be transformative in terms of both our understanding of human interactions with social media, and the ability of organizations to engage with their audience.

\section*{Acknowledgment}

The authors gratefully acknowledge both the NSF (IIS-1018361) and DARPA (YFA-N66001-12-1-4245; STTR-D13PC00064) for their support of this research.



\bibliographystyle{IEEE}
\bibliography{references}
\end{multicols*}
\end{document}